\documentclass[twocolumn,english,aps,prd,reprint,notitlepage,footinbib,preprintnumbers,superscriptaddress]{revtex4-2}

\usepackage{amsmath,amssymb,amsfonts}
\usepackage{hyperref}
\usepackage{breakurl}
\usepackage{cleveref}
\usepackage{color}
\usepackage{url}
\usepackage{lipsum}

\usepackage{graphicx}
\usepackage[export]{adjustbox}

\usepackage{accents}
\usepackage{mathrsfs}
\usepackage{mathtools}

\usepackage[usenames,dvipsnames]{xcolor}
\definecolor{skyblue}{RGB}{34,139,230}
\definecolor{navy}{rgb}{0,0,0.7}
\definecolor{purple}{RGB}{171,1,207}
\definecolor{labelcolor}{RGB}{194, 175, 116}

\definecolor{paracolor}{RGB}{0,125,255}

\usepackage[final]{showlabels} 

\def\mem{\hspace{0.1em}}
\def\hem{\hspace{0.05em}}
\def\nem{\hspace{-0.1em}}
\def\hnem{\hspace{-0.05em}}

\def\hhnem{\hspace{-0.025em}}

\def\minie{{\textstyle\frac{1}{2}}}

\def\Kerr{{\smash{\text{$\kern-0.075em\sqrt{\text{Kerr\hem}}$}}}}

\def\a{\alpha}
\def\b{\beta}
\def\c{{\gamma}}
\def\d{{\delta}}
\def\e{\epsilon}
\def\ve{\varepsilon}
\def\m{\mu}
\def\n{\nu}
\def\r{\rho}
\def\s{\sigma}

\def\bpsi{{\smash{\bar{\psi}}\kern0.02em\vphantom{\psi}}}

\def\mtimes{{\mem\times\mem}}
\def\mdot{{\mem\cdot\mem}}

\def\da{{\protect\dot{\a}}}
\def\db{{\protect\dot{\b}}}
\def\dc{{\protect\dot{\c}}}
\def\dd{{\protect\dot{\d}}}
\def\rmA{{\mathrm{A}}}
\def\rmB{{\mathrm{B}}}
\def\rmC{{\mathrm{C}}}

\renewcommand{\o}{o}

\def\bo{{\bar{o}}}

\def\lsq{{
    \kern-0.037em
    \adjustbox{scale=0.919,valign=c}{$
        {
            \adjustbox{raise=-0.0855em}{$\lfloor$}
            \llap{\reflectbox{\rotatebox[origin=c]{180}{$\lfloor$}}}
        }
    $}
    \kern-0.04em
}}
\def\rsq{{
    \kern-0.04em
    \adjustbox{scale=0.919,valign=c}{$
        {
            \rlap{\reflectbox{\rotatebox[origin=c]{180}{$\rfloor$}}}
            \adjustbox{raise=-0.0855em}{$\rfloor$}
        }
    $}
    \kern-0.037em
}}

\def\rmx{{
    \scalebox{1.2}[1]{$\mathrm{x}$}\kern-0.625em\scalebox{1.2}[1]{$\mathrm{x}$}
}}
\def\rmS{{
    \scalebox{1.09}[0.95]{$\mathrm{S}$}\kern-0.625em\scalebox{1.09}[0.95]{$\mathrm{S}$}
}}
\def\rmN{{
    \scalebox{1.2}[0.95]{$\mathrm{N}$}\kern-0.895em\scalebox{1.2}[0.95]{$\mathrm{N}$}
}}

\def\txi{{\protect\tilde{\xi}}}
\def\tzeta{{\protect\tilde{\zeta}}}

\newcommand{\wrap}[1]{{\smash{#1}\vphantom{\beta}}}

\newcommand{\dbar}{
    d\kern-.20em\makebox[0pt][l]{$\bar{}$}\kern.20em
}

\newcommand{\dinexact}{
    d\kern-.20em\makebox[0pt][l]{\scalebox{0.95}{$\bar{}$}}\kern.20em
}

\def\mtimes{{\mem\times\hem}}
\def\mdot{{\mem\cdot\mem}}

\def\vp{{\protect\vec{\partial}}}

\def\vex{\vec{x}}

\newcommand{\LOR}[1]{\smash{^\text{L}\kern-0.05em #1}}

\def\tdo{\protect\tilde{o}}
\def\ti{\protect\tilde{\iota}}

\def\tm{{\tilde{m}}}

\def\tZ{\smash{\tilde{Z}}}

\def\ram{\smash{\tilde{\pi}}}
\def\rmu{\tilde{\omega}}

\def\F{f}

\def\tchi{\tilde{\chi}}

\makeatletter
\renewcommand{\paragraph}[1]{%
    \vskip1pt%
    \@customtitle{#1}%
}
\newcommand{\intro}[1]{%
    \@customtitle{#1}%
}

\newcommand{\@customtitle}[1]{%
    {{\textit{#1}}.---}%
    {\kern0.07em}%
}
\makeatother

\newcommand{\eqs}[2]{Eqs.\,\smash{\oldeqref{#1}}\,and\,\smash{\oldeqref{#2}}}

\let\oldeqref\eqref
\renewcommand{\eqref}[1]{Eq.\,\smash{\oldeqref{#1}}}

\newcommand{\rcite}[1]{Ref.\,\cite{#1}}
\newcommand{\rrcite}[1]{Refs.\,\cite{#1}}

\usepackage{float}
\newcommand\trick[1]{}

\begin{document}

\title{
    Single Kerr-Schild Metric
    for Taub-NUT Instanton
}

\author{Joon-Hwi Kim}
\affiliation{Walter Burke Institute for Theoretical Physics, California Institute of Technology, Pasadena, CA 91125}

\begin{abstract}
    It is shown that
    a complex coordinate transformation maps the
    Taub-NUT instanton metric to a Kerr-Schild metric.
    This metric involves
    a semi-infinite line defect
    as the gravitational analog of the Dirac string,
    much like the original metric.
    Moreover, it facilitates three versions of classical double copy correspondence with the self-dual dyon in electromagnetism,
    one of which involving a nonlocal operator.
    The relevance to the Newman-Janis algorithm
    is briefly noted.
\end{abstract}

\preprint{CALT-TH 2024-020}


\bibliographystyle{utphys-modified}
\renewcommand*{\bibfont}{\fontsize{8}{8.5}\selectfont}
\setlength{\bibsep}{0.5pt}

\maketitle


\intro{Introduction}%
Recently,
an exciting program dubbed
classical double copy
\cite{monteiro2014black,luna2019type,chacon2021weyl,white2021twistorial,Luna:2022dxo,Luna:2015paa,Berman:2018hwd,elor2020newman,bah2020kerr,CarrilloGonzalez:2019gof,Carrillo-Gonzalez:2017iyj,Monteiro:2020plf,Monteiro:2021ztt,Bahjat-Abbas:2017htu,Banerjee:2019saj,Alfonsi:2020lub,monteiro2011kinematic}
has refined analogies between gravity and electromagnetism
into
concrete
mappings
between classical solutions
while being rooted in the study of scattering amplitudes
\cite{Kawai:1985xq,Bern:2010ue,Bern:2010yg,monteiro2011kinematic,Bern:2019prr}.
The prime example is the
Kerr-Schild (KS) classical double copy,
established
in the seminal work \cite{monteiro2014black}.
Namely, it is shown that
a stationary KS solution
$g = \eta + \Phi\mem \ell \otimes \ell$
in general relativity
maps to
a stationary solution in Maxwell theory
described by the gauge potential
$A = \Phi\mem \ell$,
by taking a ``single copy'' of the null one-form $\ell$.
Conversely, one envisions
a gravitational counterpart
$g = \eta + \Phi\mem \ell \otimes \ell$
of a stationary solution
$A = \Phi\mem \ell$
in electromagnetism,
by taking a ``double copy'' of $\ell$.
Well-known
instances
of this correspondence
are
Schwarzschild to point charge
and
Kerr to
a rotating solution dubbed {\Kerr}
\cite{monteiro2014black,aho2020,Lynden-Bell:2002dvr,Newman:1965tw-janis}.

From this angle,
it is natural to expect that
the Taub-NUT solution
\cite{Taub:1950ez,Newman:1963yy}
will correspond to
a dyon,
given the interpretation of the NUT charge
as the gravitomagnetic monopole
\cite{Misner:1963flatter,Bonnor:1969ala,sackfield1971physical,cho1991magnetic,dowker1967gravitational,samuel1986gravitational,maartens1998gravito,Alfonsi:2020lub}.
Unfortunately,
only a \textit{double} KS metric
$g = \eta + \Phi_1\hem \ell_1 \hnem\otimes\hnem \ell_1 + \Phi_2\mem \ell_2 \hnem\otimes\hnem \ell_2$
has been explicitly known
for Taub-NUT
\cite{Plebanski:1975xfb,Chong:2004hw},
so
a mapping to
electromagnetic dyons
had to be
achieved
by
a \textit{variant} of
the KS double copy
\cite{Luna:2015paa,Alfonsi:2020lub}.

Remarkably,
in the present letter
we point out that
the straightforward application of the \textit{original} KS double copy is
in fact
possible for the \textit{self-dual} Taub-NUT (SDTN) solution,
which is an extremal case of the Taub-NUT solution
that
has gained
a revived interest \cite{Adamo:2023fbj,Guevara:2023wlr,Crawley:2021auj,Crawley:2023brz,Adamo:2024xpc}
because of
its
intriguing physical properties
\cite{hawking1977gravitational,Gibbons:1978tef,Gibbons:1986hz,Gibbons:1987sp,dunajski2024solitons}.
The crucial fact is that
the SDTN solution remarkably admits a KS metric
in the \textit{ordinary} sense:
$g = \eta + \Phi\mem \ell \otimes \ell$.

Firstly,
a KS metric
is obtained
by KS double copying the gauge potential of
the self-dual dyon (SDD)
in electromagnetism.
We then show that
it describes the SDTN solution
by explicitly constructing a coordinate transformation to
the well-known Gibbons-Hawking
\cite{hawking1977gravitational,Gibbons:1978tef}
instanton metric.

Secondly,
the physical interpretation of the metric
as a gravitational dyon
is established
within the KS description.
Most importantly,
we confirm the characteristic
\textit{Misner string} geometry:
the gravitational analog
\cite{Misner:1963flatter,Bonnor:1969ala}
of the Dirac string
\cite{Dirac:1931kp,Dirac:1948um}.
Notably,
finding
its distributional source
\cite{Bonnor:1969ala,sackfield1971physical}
becomes a simple problem
in KS coordinates,
thanks to a linearization
\cite{Harte:2016vwo,xanthopoulos1978exact,vines2018scattering}.

Thirdly,
the
classical
double copy
is established
in two more ways.
We first show how our KS description
concretely validates the Weyl double copy
proposed by \rrcite{luna2019type,white2021twistorial,chacon2021weyl},
which has been
stated without referencing an explicit metric.
We then propose
an exotic
form of classical double copy
employing
a \textit{nonlocal} operator,
which arises from
propagators between dyonic matter
\cite{Zwanziger:1970hk,Gubarev:1998ss,Shnir:2011zz,Terning:2020dzg,Moynihan:2020gxj}.

Lastly,
we briefly
remark on
how we originally
arrived at
this KS metric,
which was, amusingly, from
the Newman-Janis shift property
\cite{Newman:1965tw-janis}
of the Kerr solution.


\paragraph{Derivation of SDTN metric from double copy}%
The gauge potential of
a dyon carrying electric and magnetic charges $(q,-iq)$
is given by
\begin{align}
\begin{split}
    \label{eq:A0}
    A
    =
    \frac{q}{4\pi r}\mem
    \Big(\hem{
        - dt
        - i\hem r\hem
        (1{\mem-\mem}\cos\theta)\mem d\phi
    }\mem\Big)
    \,,
\end{split}
\end{align}
where
we have set up spherical coordinates
and
oriented the Dirac string
\cite{Dirac:1931kp,Dirac:1948um}
along the negative $z$-axis (south pole).
This describes a complexified solution
where
the field strength $F=dA$ is self-dual:
$*F = +i\hem F$.

Notably,
adding a total derivative
$(q/4\pi)\mem d\log(r{\,+\,}z)$
gauge transforms
this
to a KS form:
\begin{align}
    \label{eq:A1}
    A
    =
    \frac{q}{4\pi r}\mem
    \bigg({
        -dt
        + \frac{x{\,+\,}iy}{r{\,+\,}z}\mem dx
        + \frac{y{\,-\,}ix}{r{\,+\,}z}\mem dy
        + dz
    }\bigg)
    \,,
\end{align}
where $x = r\sin\theta\cos\phi$, $y = r\sin\theta\sin\phi$, $z = r\cos\theta$
are the Cartesian coordinates.
Importantly,
the terms inside the bracket comprise a null one-form.

With this understanding,
the KS classical double copy \cite{monteiro2014black}
maps
\eqref{eq:A1}
to the following line element,
by literally taking ``two copies'' of the null one-form:
\begin{align}
\begin{split}
    \label{eq:g1}
    \kern-0.3em
    ds^2
    =
    {}&{}{
        -dt^2 + dx^2 + dy^2 + dz^2
    }
    \\
    {}&{}
        +
        \frac{m}{4\pi r}\mem
        \bigg({
            -dt
            + \frac{x{\,+\,}iy}{r{\,+\,}z}\mem dx
            + \frac{y{\,-\,}ix}{r{\,+\,}z}\mem dy
            + dz
        }\mem\bigg)^{\nem\hnem2}
        \,.
    \kern-0.35em
\end{split}
\end{align}
Here we have adopted $8\pi G {\,=\,} 1$ units
and replaced
the charge $q$ with the mass $m$.
Remarkably,
\eqref{eq:g1}
is a metric of
the SDTN solution,
whose mass and NUT charge are
$m$ and $-im$!
%

\paragraph{The diffeomorphism}%
To prove this,
we simply construct
an explicit coordinate transformation
to the previously known metric.
It is helpful to employ
complex coordinates
$\zeta = (x{\,+\,}iy)/(r{\,+\,}z)$,
$\smash{\tzeta} = (x{\,-\,}iy)/(r{\,+\,}z)$
that arise
from stereographically projecting
the two-sphere
with respect to the south pole,
so
\eqref{eq:g1} translates to
\begin{align}
\begin{split}
    \label{eq:g1-stereographic}
    ds^2
    =
    {}&{}{
        -dt^2 + dr^2 + r^2\mem \frac{4\hem d\smash{\tzeta}\hem d\zeta}{(1{\,+\,}\tzeta\zeta)^2}
    }
    \\
    {}&{}
        +
        \frac{m}{4\pi r}\mem
        \bigg({
            -dt
            + dr
            - r\mem \frac{2\tzeta\hem d\smash{\zeta}}{1{\,+\,}\tzeta\zeta}
        }\mem\bigg)^{\nem\hnem2}
        \,.
\end{split}
\end{align}
Amusingly, applying a diffeomorphism given by
\begin{align}
\begin{split}
    \label{eq:diff}
    t
        \,\mapsto\mem
    t
        &\,,\quad
    r
        \,\mapsto\mem
    r + \frac{m}{4\pi}
    \,,\\
    \zeta
        \,\mapsto\mem
    \zeta
        &\,,\quad
    \tzeta
        \,\mapsto\mem
    \frac{1}{
        1 + (m/4\pi r) (1{\,+\,}\tzeta\zeta)
    }\mem
    \tzeta
    \,,
\end{split}
\end{align}
we obtain
\begin{align}
    \label{eq:g0}
    ds^2
    =
    {}&{}
        -
        \bigg(\hem{
            1 + \frac{m}{4\pi r}
        }\hem\bigg)^{\nem\hnem-1}\hnem
        \bigg(\mem{
            dt
            + \frac{m}{4\pi r}\mem
            \bigg(\mem{
                dr
                - r\mem \frac{2\tzeta\hem d\smash{\zeta}}{1{\,+\,}\tzeta\zeta}
            }\mem\bigg)\nem\nem
        }\mem\bigg)^{\nem\hnem2}
    \nonumber
    \\
    &
        +
        \bigg(\hem{
            1 + \frac{m}{4\pi r}
        }\hem\bigg)
        \bigg(\mem{
            dr^2
            + r^2\hem
            \frac{4\mem d\tzeta d\zeta}{(1{\,+\,}\tzeta\zeta)^2}
        }\mem\bigg)
    \,,
\end{align}
which is the Gibbons-Hawking metric
\cite{hawking1977gravitational,Gibbons:1978tef}
for the SDTN solution
with mass $m$!

This diffeomorphism
is surely invertible.
In fact,
as is explicated in the supplemental material,
it turns out that
\eqref{eq:diff}
traces back to
the flow along the null geodesic congruence
associated with the KS metric.
Thus
the inverse map
is simply
\eqref{eq:diff}
with $m$ replaced with $-m$.

This completes our \textit{constructive} proof that
the KS and Gibbons-Hawking
metrics
in
\eqs{eq:g1}{eq:g0}
are diffeomorphic.
Consequently,
we establish that the SDTN solution admits
a construction from the KS classical double copy.
Previously,
these facts
were only recently \textit{conjectured}
by \rcite{gabriel1}
in split signature.

Meanwhile,
the astute reader may have noticed
an unusual feature of
the diffeomorphism in \eqref{eq:diff}:
it
describes a \textit{complexified diffeomorphism},
since $\zeta$ is frozen
while $\smash{\tzeta}$ gets scrambled:
hence the notation $\smash{\tzeta}$
instead of
$\smash{\bar{\zeta}}$.
Yet crucially,
this is
a legitimate operation
since
the Lorentzian-signature SDTN solution
is inherently a complex saddle
as a self-dual solution
\cite{Adamo:2023fbj}.
Just as
the SDD gauge potential
described in
\eqref{eq:A1}
is a solution to
complexified Maxwell's equations
due to the imaginary magnetic charge $-iq$,
the SDTN metric in \eqref{eq:g1}
solves
complexified Einstein's equations
governing holomorphic metrics
\cite{plebanski1975some,Plebanski:1977zz,shaviv1975general,newman1976heaven,Adamo:2023fbj}.
As a consequence
of this interpretation,
$\zeta$ and $\smash{\tzeta}$ can be transformed independently as \textit{holomorphic} coordinates.

\paragraph{Physical interpretation}%
We shall now examine the physical validity of the
KS metric in \eqref{eq:g1}.
To begin with,
we point out that
\eqref{eq:g1}
\textit{itself}
is quite unusual
as a KS metric,
in fact.
While the KS potential is simply
$\Phi = m/4\pi r$,
the null one-form
$\ell$
involves a \textit{semi-infinite line defect} along the negative $z$-axis,
which
is exactly how
the gravitomagnetic charge
is encoded
in this metric:
\begin{align}
    \label{eq:ell-zag}
    \ell
    \mem=
        -dt
        + \frac{x{\,+\,}iy}{r{\,+\,}z}\mem dx
        + \frac{y{\,-\,}ix}{r{\,+\,}z}\mem dy
        + dz
    \,.
\end{align}
This is intriguing,
as the null one-form is usually regular
wherever the KS potential is regular---%
which is especially true for
the double KS construction of Taub-NUT \cite{Luna:2015paa}.

Crucially,
as in the original metric in \eqref{eq:g0},
the interpretation
of this line defect
should be the \textit{Misner string}
\cite{Misner:1963flatter,Bonnor:1969ala}:
the gravitational analog of the Dirac string,
inputting a gravitomagnetic flux of $-im$
at the origin.
To show this, we recall that
there are at least three senses in which
the Taub-NUT metric
\cite{Misner:1963flatter,Bonnor:1969ala}
describes
a gravitational analog of a dyon:
(a)
    the Misner string is
    invisible as
    a coordinate artifact
    if
    the gravitational analog of the Dirac quantization condition holds \cite{Misner:1963flatter},
(b)
    the Misner string implements
    a flux tube of
    time monodromy
    \cite{cho1991magnetic,dowker1967gravitational,Alfonsi:2020lub},
and
(c)
    the distributional stress-energy of the string
    describes
    an energy-less thin solenoid of mass current \cite{Bonnor:1969ala,sackfield1971physical}.
Our goal now is to
show that all of these properties
hold
in the KS description as well.

Firstly, we derive the Dirac quantization condition.
For the SDD gauge potential in \eqref{eq:A1},
adding a term
\smash{$(q/2\pi)\mem d\log\zeta$}
repositions its Dirac string
along the positive $z$-axis (north pole).
This encodes
the complexified group-valued
transition function
\smash{$\zeta{}^{\hem ieq/2\pi}$},
where $e$ is the electromagnetic coupling.
Demanding its single-valuedness
on the ``equator''
\cite{Wu:1975es}
implies
$ieq \in 2\pi\mathbb{Z}$.

The gravitational Dirac quantization
can be derived
in a similar fashion,
i.e., by
hiding the strings with overlapping coordinate patches
\cite{Misner:1963flatter,dowker1967gravitational,cho1991magnetic,Alfonsi:2020lub}.
In the KS description,
we find that
the diffeomorphism
\begin{align}
\begin{split}
    \label{eq:reposition}
    t
        \,\mapsto\mem
    t
    + \frac{m}{2\pi} \log \zeta
    \,,\,\,\,\,
    \tzeta
        \,\mapsto\mem
    \frac{
        1 - (m/4\pi r) (1{\,+\,}\tzeta\zeta)/\tzeta\zeta
    }{
        1 + (m/4\pi r) (1{\,+\,}\tzeta\zeta)
    }\mem \tzeta
\end{split}
\end{align}
with $r$ and $\zeta$ fixed
transforms
the KS metric in
\eqref{eq:g1-stereographic} to
another KS metric:
\begin{align}
\begin{split}
    \label{eq:north-g1-stereographic}
    ds^2
    =
    {}&{}{
        -dt^2 + dr^2 + r^2\mem \frac{4\hem d\smash{\tzeta}\hem d\zeta}{(1{\,+\,}\tzeta\zeta)^2}
    }
    \\
    {}&{}
        +
        \frac{m}{4\pi r}\mem
        \bigg({
            -dt
            + dr
            + r\mem \frac{2\hem d\smash{\zeta}}{\zeta(1{\,+\,}\tzeta\zeta)}
        }\mem\bigg)^{\nem\hnem2}
        \,,
\end{split}
\end{align}
which develops a string defect along the positive $z$-axis
(north pole).
For this diffeomorphism
to be single-valued within the overlapping region,
the time coordinate
should be periodic as
$t \sim t + im \mathbb{Z}$,
given the $2\pi i$ periodicity of $\log \zeta$.
Hence
the energy $e$
of a test particle
is quantized as
$iem \in 2\pi\mathbb{Z}$.
This is precisely the gravitational
Dirac quantization condition
due to the NUT charge $-im$.


Secondly, we study the near-string geometries
in an explicit manner.
Let
$\xi := x+iy$,
$\smash{\txi} := x-iy$
be complex coordinates
for the (complexified) $x$-$y$ plane,
and suppose a positive infinitesimal $\e$.
Applying the replacement $z \mapsto -1/\e + z$
to the gauge potential in \eqref{eq:A1}
and then taking the limit $\e \to 0$,
we obtain
$A = (q/2\pi)\mem d\log\smash{\txi}$.
This precisely describes an infinite Dirac string
generated by a large gauge transformation
\cite{tong2018gauge,Cheung:2024ypq}.

For the line element in
\eqref{eq:g1},
the same limiting procedure
leaves out a divergent part
that
can be compensated by
a shift
$\smash{\xi} \mapsto \smash{\xi} + (
    (m/\pi)\mem (1/\e{\,-\,}z)
    - 3m^2 \nem/4\pi^2
) /\smash{\txi}$.
Amusingly,
the resulting line element is
simply the Minkowski line element,
but with a multivalued time:
\begin{align}
    t' = t + \frac{m}{2\pi}\mem \log\txi
    \,.
\end{align}
Namely,
in the vicinity of the string,
the geometry approaches to
the image of empty spacetime
under
a large diffeomorphism.
Here,
a Sagnac interferometer
measures a monodromy
in time as
the gravitational Aharanov-Bohm phase
\cite{sakurai1980comments,dowker1967gravitational,samuel1986gravitational,zimmerman1989geodesics},
localized along the $z$-axis (as $\smash{\txi} = 0$).
This precisely describes
an infinite Misner string
as a thin tube of gravitomagnetic flux $-im$
\footnote{
    One might make a parallel between the field strength
    $F = dA$ and the torsion two-form $ddt'$
    (cf.\,\rrcite{Fiziev:1995te,Kleinert:1996yi,Kleinert:2008zzb})
    associated with these two strings,
    which are distributional.
}.

\begin{figure}[t]
    \centering
        \includegraphics[valign=c,width=0.55\linewidth]{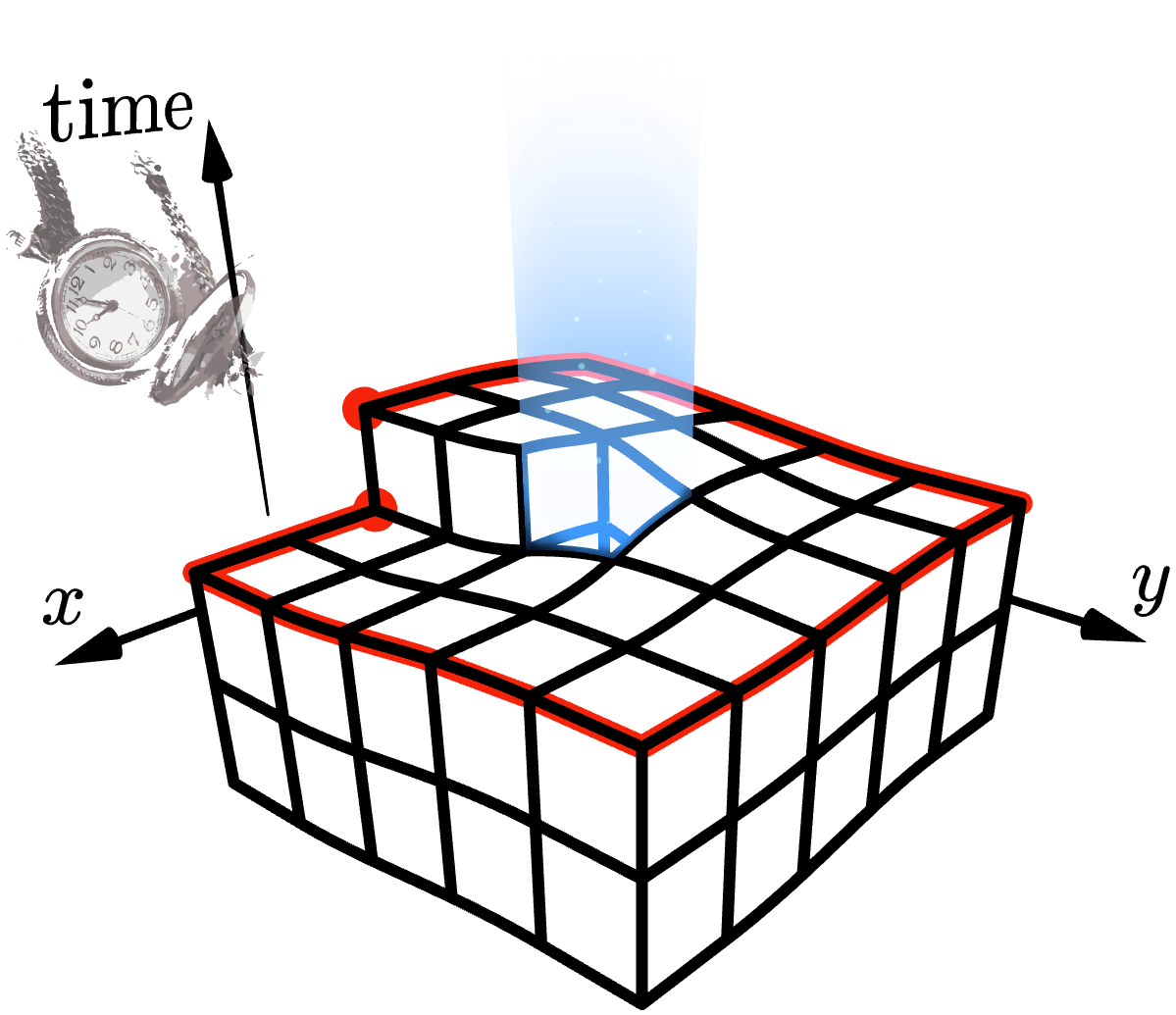}
    \caption{
        The Misner string is a ``screw dislocation''
        in spacetime,
        inducing time monodromy as an Aharanov-Bohm phase.
    }
    \label{fig:misner}
\end{figure}

Thirdly,
we derive the distributional stress-energy tensor
of this string.
Among various approaches
\cite{israel1970source,Balasin:1993kf,Balasin:1993fn},
we particularly
consider
the ``harmonic-gauge linearization''
of \rcite{vines2018scattering}
which
stationary KS metrics enjoy
as a luxury.
Provided
the congruence of $\ell^\m\mem \partial_\m$
is
geodesic and shear-free (GSF),
this approach
identifies
\smash{$T^{\m\n} {\,=\,} u^{(\m} J^{\n)}$}
as the source
for the double copy solution
\cite{vines2018scattering}.
Here
$J^\m$ is the single copy source,
while $u^\m {\,=\,} \delta^\m{}_0$
is the unit timelike Killing vector.
For our case,
the GSF property of the congruence associated with the null one-form in \eqref{eq:ell-zag}
can be established from an identity
\begin{align}
    \partial_\m \ell_\n
    =
    \frac{1}{2r}\mem
    \Big(\,{
        \eta_{\m\n}
        + \ell_\wrap{(\m} (2u{\,-\,}\ell)_\wrap{\n)}
        - i\mem \ve_{\m\n\r\s} u^\r \ell^\s
    }\mem\Big)
    \,,
\end{align}
while
the single copy source $J^\m$
easily follows by
acting on the Laplacian $\square := -\eta^{\m\n}\partial_\m\partial_{\n}$
to \eqref{eq:A0}.
Applying the formula
\smash{$T^{\m\n} {\,=\,} u^{(\m} J^{\n)}$}
then immediately reproduces
Bonnor's stress-energy tensor
\cite{Bonnor:1969ala,sackfield1971physical},
describing a point mass $m$
attached to
a massless
semi-infinite
rotating rod.

To sum up,
we have convinced ourselves
from multiple angles
that the KS metric in \eqref{eq:g1} indeed
exhibits the geometrical and physical
characteristics of
a Taub-NUT spacetime
arising from the Misner string.
This analysis
demonstrates
that
the KS description of the solution
is not merely a formal construct
but physical.
Moreover,
the quick derivation of the distributional source
demonstrates its usefulness
as well.


\paragraph{Weyl double copy in position space}%
Finally,
let us
unlock more classical double copy relations
for the SDTN solution.
Firstly,
we explicate
how our KS metric
solidly verifies
the \textit{Weyl double copy} correspondence
\cite{chacon2021weyl,white2021twistorial}.

The Weyl double copy
directly constructs the Weyl curvatures
of algebraically special spacetimes
by ``squaring'' electromagnetic field strengths
\cite{luna2019type,chacon2021weyl,white2021twistorial,Luna:2022dxo}
while not
referencing
a metric nor a gauge potential.
For example,
take the Coulomb and Schwarzschild solutions.
In the spinor language,
their field strength and Weyl tensor are
\begin{align}
    \label{eq:NPformalism}
    F_\wrap{\da\db}
    = \frac{2q}{4\pi r^2}\mem
        \tdo_\wrap{(\da} \ti_\wrap{\db)}
    \,,\,\,\,\,
    C_\wrap{\da\db\dc\dd}
    = -\frac{6m}{4\pi r^3}\mem
        \tdo_\wrap{(\da} \tdo_\wrap{\db}
        \hem
        \ti_\wrap{\dc} \ti_\wrap{\dd)}
    \,,
\end{align}
where we have omitted the anti-self-dual counterparts.
The Weyl double copy is
the squaring relation
$C_\wrap{\da\db\dc\dd} = F_\wrap{(\da\db} F_\wrap{\dc\dd)} / S$
\cite{luna2019type},
which holds
with $S {\,\propto\,} 1/r$.

Crucially, this relation is \textit{chiral}:
there is no mixing between the self-dual and anti-self-dual parts.
Thus the Weyl double copy for SDTN
seems to simply follow from
that of Schwarzschild,
if one posits that
the SDTN solution
is the ``self-dual part'' of the Schwarzschild solution
in the sense that
its self-dual
Weyl curvature
and the principal spinors
exactly equal those of the Schwarzschild solution
\cite{chacon2021weyl,white2021twistorial}.
This physical anticipation, however,
may appear a bit subtle
due to the nonlinearity of gravity.
However luckily,
straightforward calculations with
the explicit KS metric
in \eqref{eq:g1}
concretely verifies this claim,
so
the Weyl double copy construction
of the SDTN solution
is completely validated
at the fully nonlinear order.

Specifically,
the principal spinors
for the KS metric in \eqref{eq:g1}
are found as
$\tdo^\da = (1,\zeta)$
and
$\ti^\da {\,=\,} ( -\smash{\tzeta} , 1 ) /(1{\,+\,}\smash{\tzeta}\zeta)$,
one of which being encoded in the null one-form:
\begin{align}
    \label{eq:lada}
    \ell^\m
    = (\s^\m)_{\a\da}\mem
        \tdo^\da \o^\a
    \,,\quad
    \tdo^\da = \begin{pmatrix}
        1 \\ \zeta
    \end{pmatrix}
    \,,\quad
    \o^\a = \begin{pmatrix}
        1 \\ 0
    \end{pmatrix}
    \,.
\end{align}
Crucially,
the \textit{real} combinations
$[\tdo_\da]^*\hem \tdo_\db$,
$[\ti_\da]^*\hem \ti_\db$
of these spinors
exactly reproduce
the null one-forms $-dt \pm dr$
associated with the
KS metrics of the
Schwarzschild solution
as stated in
\rrcite{white2021twistorial,chacon2021weyl}
\footnote{
    This is consistent with the fact that
    the GSF spinor equation
    \cite{penrose1984spinors2,huggett1994introduction}
    is chiral
    (cf. \rcite{gabriel1}).
}.

\paragraph{Weyl double copy in twistor space}%
Meanwhile, the Weyl double copy has been also discussed in \textit{twistor space} \cite{white2021twistorial,chacon2021weyl}.
This is essentially a statement about
the \textit{linearized} Weyl tensor
as a massless spin-$2$ field
on flat background
constructed from Penrose transform
\cite{penrose1984spinors2,Penrose:1969ae,Atiyah:2017erd}.
The Penrose transform for a massless spin-$h$ field is given by
\footnote{
    Note that
    \smash{$(\F(\protect\tZ))^{-h-1}$}
    could be interpreted as
    the half-Fourier transformed
    scattering amplitude
    of a massive object
    absorbing a negative-helicity force carrier
    \cite{guevara2021reconstructing}.
}
\begin{align}
    \label{eq:Penrose}
    \phi_{\da_1\cdots\da_{2h}}\nem(x)
    \mem=
    \oint \frac{
        \ram_\dc\hem d\ram^\dc
    }{2\pi i}\,
        \frac{
            \ram_{\da_1} \nem\cdots \ram_{\da_{2h}}
        }{
            ({ \F(\ram,\ram x) }\hhnem)^{h+1}
            \vphantom{\tilde{Z}}
        }
    \,,
\end{align}
and
the self-dual linearized Weyl tensor of the Schwarzschild and SDTN solutions
arises from \eqref{eq:Penrose}
for $h=2$
by taking the twistor function
\cite{penrose-TN1-10,sparling-TN1-14,hughston1979advances,white2021twistorial}
\footnote{
    This function
    can be written as
    $i\mem
    \protect\tZ^\rmA\mem
    u_\m (\gamma^\m)_\rmA{}^\rmC\mem
    \protect\bar{I}_{\rmC\rmB}\mem
    \protect\tZ^\rmB$
    by employing the infinity twistor
    $\protect\bar{I}_{\rmA\rmB}$.
}
\begin{align}
    \label{eq:F}
    \F(\tZ)
    =
        -
        \rmu^\a\mem u_{\a\da}\mem \ram^\da
    \,,\quad
    u_{\a\da} := u_\m (\s^\m)_{\a\da}
    \,,
\end{align}
whose zeros
encode the principal spinors
$\tdo_\da,\ti_\da$
as per
the Kerr theorem
\cite{penrose1967twistoralgebra,penrose1984spinors2,newman2004maxwell}.
The twistor-space Weyl double copy
implies that
taking $\smash{h=1}$ with the same function
derives the single copy field strength
in \eqref{eq:NPformalism}
\cite{white2021twistorial,chacon2021weyl}.

Here,
the crucial clarification which
the explicit KS metric
provides
is that
this twistor-space construction
from Penrose transform
has
in fact
described the \textit{exact} Weyl tensor of the SDTN solution
at the fully nonlinear level.
Namely,
the Cartesian coordinates of the flat background employed in the Penrose transform
secretly described KS coordinates for the curved SDTN geometry.

Note that
the ``position vector'' $\vex$
which one identifies from the Killing spinor
as
\smash{$
    \tchi^\da{}_\db
    = \minie\mem (\vex {\mem\cdot\mem} \vec{\sigma})^\da{}_\db
$}
\cite{Adamo:2023fbj}
also describes the (spatial part of) our KS coordinates.

\paragraph{Self-dual double copy}%
Moreover,
our KS coordinates are also
the Pleba\'nski coordinates
\cite{plebanski1975some,Adamo:2021bej}
for the second heavenly equation,
on which the
\textit{self-dual double copy}
\cite{monteiro2011kinematic,monteiro2014black,Berman:2018hwd}
is based.
Notably,
the self-dual double copy
would imply
fascinating conclusions such as
``SDTN \textit{is} a large-$N$ dyon''
in the context of
Moyal deformation of the heavenly equation
\cite{Plebanski:1994qi,Park:1990fp,Park:1990vi,Husain:1993dp,Cheung:2022mix,Armstrong-Williams:2022apo}.
Strangely,
however,
the SDTN metric
is not straightforwardly
the self-dual double copy
of SDD
in terms of the Pleba\'nski scalar field
\footnote{
    It might be that
    the color-to-diffeomorphism replacement rule
    \cite{Cheung:2021zvb}
    directly applies
    at the level of the ``chiral current''
    so that
    $A_\wrap{1\da} \sim \tdo_\da$
    maps to $h_\wrap{1\da1\db} \sim \tdo_\da \tdo_\db$.
}:
$A_{\a\da} = \o_\a \partial_{0\da}
(\hem{
    q\zeta/4\pi
}\hem)$,
$h_\wrap{\a\da\b\db}
=
    \o_\a \o_\b\mem \partial_\wrap{0\da} \partial_\wrap{0\db}
    $ $
    (\hem{
        m (2r{\,+\,}z)\mem \zeta^2
        \hnem/24\pi
    }\hem)
$,
if $h_{\m\n} = g_{\m\n} {\,-\,} \eta_{\m\n}$
denotes the KS metric perturbation.
This puzzle deserves further investigation,
in comparison with the
Eguchi-Hanson case \cite{Berman:2018hwd}.

\paragraph{Operator KS double copy}%
Interestingly,
we still find that an operator KS double copy
is viable,
but in a more exotic form.
The operator KS double copy
promotes
the null one-form $\ell_\m$
in the KS ansatz
to a differential operator $\smash{\hat{k}_\m}$,
which
may factorize as
$-\minie\mem (\s^\m)_{\a\da}\mem \smash{\hat{k}_\m} = \o_\a \smash{\hat{k}_\da}$
so that the fields are
manifestly
put in the lightcone gauge
\cite{monteiro2014black,Berman:2018hwd,elor2020newman}.
Intriguingly,
we find that
\eqs{eq:A1}{eq:g1}
precisely
arise in this way
if
the operator
\smash{$\hat{k}_\da$} is given by
\begin{align}
    \label{eq:khat}
    \hat{k}_\da
    \mem=\hem
    \bigg(\mem{
        \hhnem-
        \frac{
            \partial_x {\mem+\,} i\partial_y
        }{
            \partial_z
        }
        \mem,\mem
        1
    }\mem\bigg)
    \,,
\end{align}
which is \textit{nonlocal}.
Specifically,
it follows from
$\smash{\hat{k}_\da} (1/r) = (1/r)\mem \tdo_\da$
and
$\smash{\hat{k}_\da}\smash{\hat{k}_\db} (1/r)
= \smash{(1/r)\mem \tdo_\da \tdo_\db}$
that
a double copy correspondence
between
SDD
and SDTN
is realized as
\begin{align}
    \label{eq:operator-dc}
    A_{\a\da} =
    \o_\a \smash{\hat{k}_\da}
    \mem
        \frac{q}{4\pi r}
    \,,\quad
    h_\wrap{\a\da\b\db} =
    \o_\a\o_\b\mem \smash{\hat{k}_\da}\smash{\hat{k}_\db}
    \mem
        \frac{m}{4\pi r}
    \,.
\end{align}

Notably,
this new form of classical double copy
admits a nice physical explanation
from an electric-magnetic dual perturbation theory.
The key fact is that
the SDD can be viewed as an object in the ``extended'' Maxwell theory
where the violation of Bianchi identity
by \textit{magnetic matter} $J^\star$
is allowed by the introduction of the dual gauge potential $A^\star$
\cite{Dirac:1948um}:
$F = dA + \ast dA^\star$.
In this picture, the distributional source for the SDD is
\begin{align}
\begin{split}
    \label{eq:Jdual}
    J^\m
    =
        \partial_\n F^{\m\n}
    &=
    qu^\m\mem\delta^{(3)}\hnem(\vex)
    \,,\\
    J^{\star \m}
    =
        -\partial_\n {*}F^{\m\n}
    &=
    -iqu^\m\mem\delta^{(3)}\hnem(\vex)
    \,.
\end{split}
\end{align}
We may redescribe these in the ``helicity basis'' as
$J_+^\m = J^\m - i\mem J^\star{}^\m = 0$
and
$J_-^\m = J^\m + i\mem J^\star{}^\m = 2qu^\m\mem \smash{\delta^{(3)}(\vex)}$,
while having
\smash{$A^\pm_\m := \minie\mem (A^{\vphantom{+}}_\m {\,\pm\,} iA^\star_\m)$}
for the potentials.

As a dynamical field theory,
a formulation of this extended Maxwell theory
has been given by Zwanziger
\cite{Zwanziger:1970hk,Gubarev:1998ss,Shnir:2011zz,Terning:2020dzg,Moynihan:2020gxj}.
We note that
the propagators of this theory
can be summarized into
a single formula in the spinor notation:
\begin{align}
\begin{split}
   \label{eq:zprop}
   \hat{\Delta}^{+-}_{\wrap{\a\da\b\db}}
   =
   \frac{
       n_{\wrap{\a\db}}\hem \partial_{\wrap{\b\da}}
   }{-n\mdot \partial}
   \frac{1}{\square}
   \,,
\end{split}
\end{align}
where $n$ is an auxiliary four-vector
that traces back to the Dirac string.
\eqref{eq:zprop}
describes the position-space propagator
from \smash{$A^-_\wrap{\b\db}$} to \smash{$A^+_\wrap{\a\da}$},
while the $\langle A^- A^- \rangle$ and $\langle A^+ A^+ \rangle$
propagators vanish.
A simple calculation then shows that the SDD gauge potential in \eqref{eq:A1}
is precisely reproduced by this propagator,
if one takes
$n_\wrap{\a\da} = { -\o_\a \bo_\da }$
for the auxiliary vector
so that
$\smash{\hat{k}}_\da = {- \partial_\wrap{1\da} / \partial_\wrap{0\dot{0}} }$.

Amusingly,
the graviton field
arises in the same fashion by
using the linearized gravity equivalent of Zwanziger propagator,
which is given by doubling \eqref{eq:zprop}
as is advocated in \rcite{Moynihan:2020gxj}.
In this case,
the sources in the two-potential theory
are
$T_{+}^{\m\n} = 0$ and
$T_{-}^{\m\n} = 2m u^{\m} u^{\n}\mem \smash{\delta^{(3)}\hnem(\vex)}$.
The resulting linearized metric perturbation
happens to be
also the full metric perturbation
by taking the KS form
\cite{Harte:2016vwo,xanthopoulos1978exact,vines2018scattering}.

In summary,
the double copy structure proposed in \eqref{eq:operator-dc}
can be traced back to
propagators between dyonic matter
in
electrodynamics and linearized gravity:
\begin{align}
    A^+_\wrap{\a\da}
    =
        \hat{\Delta}^{+-}_\wrap{\a\da\b\db}\mem
            J_-^{\db\b}\mem
    ,\,\,\,\,
    h^+_\wrap{\a\da\b\db}
    =
        \hat{\Delta}^{+-}_\wrap{\a\da\c\dc}\hem
        \hat{\Delta}^{+-}_\wrap{\b\db\d\dd}\mem
            T_-^{\dc\c\dd\d}\mem
    .
\end{align}
A few comments are in order.
Firstly,
it is instructive to translate the propagator in \eqref{eq:zprop}
into the vector notation,
where one can identify a combination of operators
\smash{$(\vec{n}\mtimes\vp/\vec{n}\mdot\vp)$}
connecting between the Gilbertian and Amp\`erian implementations \cite{griffiths3ed}
of multipoles.
This implies that the nonlocality of
the operator in \eqref{eq:khat}
has encoded the Dirac string structures.
Secondly,
our construction
could be justified
from the on-shell perspective.
The ``numerator'' of
the off-shell propagator in \eqref{eq:zprop}
correctly factorizes into the null polarization vectors
so that
the double copy of sources from
$\smash{J_-^\m} = \smash{2qu^\m\mem \delta^{(3)}\hnem(\vex)}$
to
$\smash{T_-^{\m\n}} = \smash{2mu^\m u^\n\mem \delta^{(3)}\hnem(\vex)}$
encodes
the squaring relation of the massive on-shell amplitudes
\cite{Huang:2019cja,Emond:2020lwi,ahh2017,Moynihan:2020gxj}.
In this perspective,
the nonlocality traces back to that of the $x$-factor.
Lastly,
it is interesting to
apply this approach to construct
generic Taub-NUT spacetimes,
in which case one naturally obtains a double KS metric perturbation.

\paragraph{Conclusions}%
In this work, we showed that
a complexified diffeomorphism
reveals
a KS metric for the SDTN solution.
The validity of this diffeomorphism and the metric was
then verified
while highlighting their complex nature.
Especially,
the KS metric
also
exhibits the Misner string defect,
supporting its physical interpretation as a gravitational dyon.
Further,
we established
three concrete senses
in which
the SDTN solution
is precisely the gravitational counterpart of
the SDD in electromagnetism:
KS,
Weyl,
and non-local operator KS
classical double copies.
Notably,
the last one
adds to
the inventory of
classical double copies
as a new entry,
whose physical origin
is the
propagators
between dyonic mater.

The most distinctive charm of this work
lies in the exciting development
along the very first three equations.
The usual punchline of
classical double copy
has been
starting with an gravitational solution
and then
deducing the electromagnetic counterpart.
This work,
in contrast,
takes a reverse path:
we constructed an electromagnetic solution
and then applied the double copy to yield a new statement in general relativity.

It is also insightful to frame this work
in the angle of
``classical double copy of gravitational instantons.''
The only instance has been
the self-dual double copy for
the Eguchi-Hanson instanton
\cite{Eguchi:1978xp,Eguchi:1978gw,Eguchi:1979yx}
provided by the work \cite{Berman:2018hwd},
which traces back to
\rrcite{Tod:1982mmp,Sparling:1981nk}.
This work
adds another instance to the discourse,
namely
the Taub-NUT instanton.
It is amusing that
the correspondence between
self-dual objects in electromagnetism and gravity
which the Gibbons-Hawking
ansatz \cite{hawking1977gravitational,Gibbons:1978tef}
had suggested a long time ago
has been strengthened from new perspectives
by the advent of the classical double copy program.

The KS metric
of the SDTN solution
conveys
an aesthetic elegance
in terms of
its
minimalist
and
spinorial nature.
To investigate
what
facilitates
this remarkable
gem,
we could
examine
its relation to
the self-dual limit of
the \textit{double} KS metric
in \rcite{Luna:2015paa}
\footnote{
    Finding a diffeomorphism bridging between the two
    would
    shed an insight.
    We thank Andres Luna for suggesting this direction.
}.

Lastly,
we anticipate that
the explicit KS metric
we have established here
will contribute to
future studies on the SDTN solution,
both conceptually and practically.
In fact,
we have already seen that
it
largely
clarifies the subtle aspects of
the Weyl double copy.
For further consequences or applications,
we suggest the following directions:
    asymptotic symmetries \cite{Godazgar:2019dkh,Godazgar:2019ikr}
    \footnote{
        The KS metric
        nicely falls into the
        BMS
        category of asymptotically flat metrics \cite{Bondi:1962px,Sachs:1962wk,Strominger:2017zoo},
        with
        mass aspect $m$,
        angular momentum aspects
        zero,
        and
        $\smash{C_\wrap{\zeta\zeta}} =$ $ \smash{4M \tzeta{}^2 \nem/ (1+\tzeta\zeta)^2}$,
        $\smash{C_{\tzeta\tzeta}} = 0$.
    },
    integrability of
    test-particle motion
    \cite{Gibbons:1986hz,Gibbons:1987sp},
    black hole
    perturbation theory
    \cite{Guevara:2023wlr,Adamo:2023fbj,Teukolsky:1973ha,Press:1973zz},
    or
    exact graviton propagators
    \cite{Sasaki:2003xr,Mano:1996vt,Bonelli:2021uvf},
    which will facilitate self-force calculations
    \cite{Poisson:2011nh,DeWitt:1960fc,Barack:1999wf,Cheung:2023lnj}
    in an idealized set-up.
Especially,
explorations in the
scattering context
are likely to
shed insights on
the Kerr black hole scattering
for an integrable subsector,
on account of
the Newman-Janis shift
\cite{Newman:1965tw-janis}:
the ``self-dual part'' of Kerr
is a SDTN solution
\cite{Newman:1973yu,Newman:2002mk,Ghezelbash:2007kw,Crawley:2021auj,gabriel1,Adamo:2023fbj}.

\begin{figure}[t]
    {\,\includegraphics[valign=c,width=0.98\linewidth]{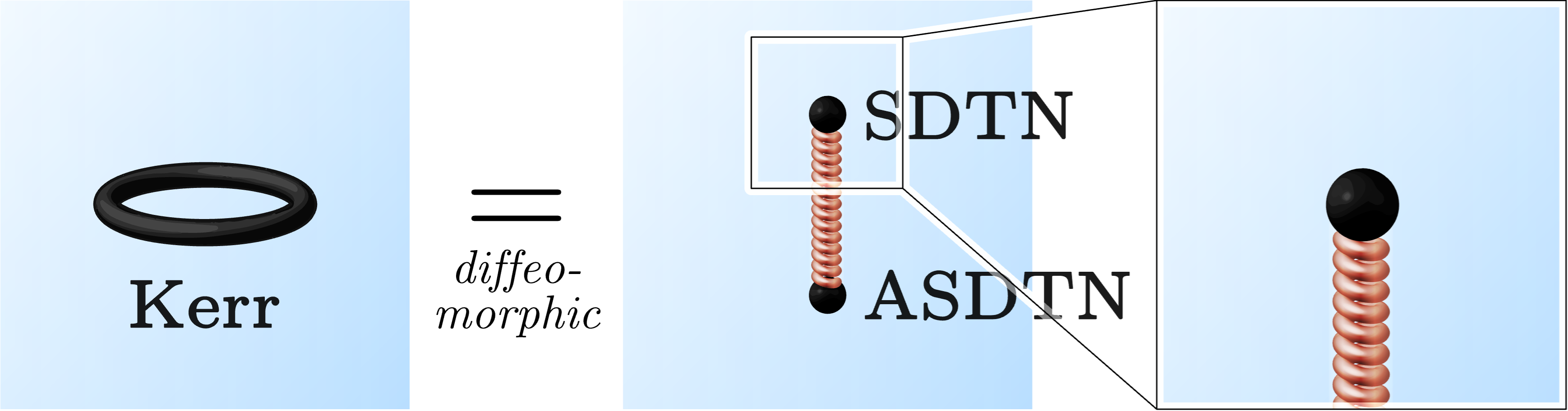}\,}
    \caption{
        The very physical origin of the Newman-Janis shift is
        the fact that
        the Kerr black hole is secretly
        a ``dumbbell'' of
        Taub-NUT instantons.
    }
    \label{fig:nj}
\end{figure}

Amusingly,
about a year and a half ago,
we deciphered
a complete physical picture
behind the Newman-Janis algorithm,
generalizing
the above statement.
As is depicted in Fig.\,\ref{fig:nj},
we realized that
a simple complex coordinate transformation
maps the Kerr metric
in KS coordinates to
another KS metric describing
two point masses joined by a Misner string.
By zooming to the upper tip of the string,
we \textit{by construction} obtained
a KS metric,
which was precisely \eqref{eq:g1}.
Crucially,
this had to be describing a Taub-NUT instanton
on account of the Newman-Janis shift!
The effort to verify this strong physical conviction
gave birth to this work.
We will release more details in the future.

\paragraph{Note added}%
The work \cite{gabriel2},
which appeared a week after the initial release of this paper,
provided an \textit{existence} proof for a real diffeomorphism
connecting between the KS and Gibbons-Hawking metrics
in split signature.
However, this paper has already provided a \textit{constructive} proof
for all signatures
by taking the complexified point of view;
e.g., the result and conjecture of \rrcite{gabriel2,gabriel1}
are reproduced by taking the real slice
$t,x,iy,z \in \mathbb{R}$.


\medskip\noindent\textit{Acknowledgements}.\:
We
would like to thank
{
    Tim Adamo%
,
    Clifford Cheung%
,
    Maciej Dunajski%
,
    Jung-Wook Kim%
,
    Andres Luna%
,
    Donal O'Connell%
,
    Lionel Mason%
,
    Nabha Shah%
,
    Justin Vines%
, and
    Chirs D. White%
}
for discussions and insightful comments.
We are also grateful to
Gabriel Herczeg, Max Pezzelle, and Jash Desai
for bringing the author's attention to the works
\cite{gabriel1,gabriel2}
after the initial release of this paper
and also the follow-up discussions.
This material is based upon work supported by the U.S. Department of Energy, Office of Science, Office of High Energy Physics, under Award Number DE-SC0011632.
J.-H.\,K. is also supported by Ilju
Academy and Culture Foundation.


\section*{
    Supplemental Material:\\
    Kerr-Schild to Gibbons-Hawking\\
    by null geodesic flow
}

Suppose a vector field $\ell^\m$ in complexified Minkowski space
that defines a geodesic congruence:
$\ell^\m{}_{,\n}\mem \ell^\n = 0$.
The time-$\e$ flow generated by this vector field
is simply
$x^\m \mapsto
x^\m + \e\mem \ell^\m(x)
$,
as
geodesics are straight lines in flat space.
The pull-back of the flat metric
$\eta =
    \eta_{\m\n}\mem dx^\m {\mem\otimes\mem} dx^\n$
is
$
    \eta +
        2\e\mem \ell_\wrap{(\m,\n)}\mem
    dx^\m {\mem\otimes\mem} dx^\n
        + \e^2\mem \eta_{\r\s}\mem \ell^\r{}_{,\m}\mem \ell^\s{}_{,\n}
    dx^\m {\mem\otimes\mem} dx^\n
$.

The null vector field associated with the KS metric of the SDTN solution,
$\ell^\m = (1 , \zeta, -i\zeta , 1)$,
is stationary, geodesic, and shear-free.
With the self-duality condition,
they together implies that
$2\hem \partial_\n \ell^\m = \rho\mem \tilde{m}^\m m_\n$,
when described by a null tetrad $(\ell^\m,n^\m = 2u^\m {\mem-\,} \ell^\m,m^\m,\tm^\m)$
where $u^\m := \delta^\m{}_0$.
The complex expansion computes to $\rho = 1/r$.

As a result,
it follows that
the flat metric transforms to
$\eta + \e\mem \rho\mem \tm {\mem\odot\mem} m$
if the geodesic flow $x^\m \mapsto x^\m + \e\mem \ell^\m(x)$
is used
as a diffeomorphism,
where $\odot$ denotes the symmetrized tensor product.
In turn,
the KS metric
$\eta + \phi\mem \ell^2$
transforms to
\begin{align}
    \eta \,+\, \e\mem \rho\mem \tm {\mem\odot\mem} m
    \,+\, \phi'\mem \ell^2
    \,,
\end{align}
where $\phi'(x) := \phi(\mem x {\,+\,} \e\mem \ell(x))$.

Meanwhile,
using $\eta = -2u {\mem\odot\mem} \ell + \ell^2
+ \tm {\mem\odot\mem} m$,
the Gibbons-Hawking metric
$-(1{\,+\,}\phi)^{-1}\mem
(\hem{
    u - \phi\mem (\ell{\,-\,}u)
}\hem)^2
+ (1{\,+\,}\phi)\mem (u^2 {\,+\,} \eta)$
boils down to
\begin{align}
\begin{split}
        \eta \,+\, \phi\mem \tm {\mem\odot\mem} m
        \,+\, (\phi^{-1} {\,+\,} 1)^{-1}\mem \ell^2
    \,.
\end{split}
\end{align}
Therefore, if we can flow along the geodesic congruence
such that $\phi^{-1} {\,+\,} 1 = \phi'^{-1}$
and $\e\mem \rho = \phi$,
then the KS metric
is mapped to the Gibbons-Hawking metric.
Evidently, $\e = m/4\pi$
satisfies both of these conditions.
Thus the diffeomorphism from the KS to Gibbons-Hawking metrics
is given by
$t \mapsto t + m/4\pi$,
$x \mapsto x + m\zeta/4\pi$,
$y \mapsto y - im\zeta/4\pi$,
$z \mapsto z + m/4\pi$,
which is equivalent to \eqref{eq:diff}
up to dropping the constant shift in the time coordinate.
It could be interesting if this ``trick''
also proves to be
useful for more general cases
such as
non-self-dual solutions
or multi-centered Gibbons-Hawking metrics.


\bibliography{references.bib}

\end{document}